\documentstyle[prd,aps,epsfig]{revtex}

\begin{document}

\title{\bf The flux-tube phase transition }
\author{G.A. Kozlov, G. Khoriauli\\
Joint Institute for Nuclear Research,\\
  Joliot-Curie st. 6, Dubna, 141980 Moscow Region, Russia}

\date{\today}
\maketitle

\begin{abstract}

 We consider the phase transition in the "hot" dual
long-distances Yang-Mills theory
at finite temperature $T$. This phase transition is associated with
a change of symmetry. The dual model is formulated in terms of two-point
Wightman functions with equations of motion involving higher derivatives.
The effective mass of the dual gauge field
is derived as a function of $T$-dependent gauge coupling constant.

\end{abstract}

\vspace{1 cm}

\section{Introduction}

Most expositions of dual model focus on its possible use as a framework
for quark confinement in nature. Rather than the confinement of color
charges, we will here describe what one might call the phase transition
in the dual Yang-Mills (Y-M) theory at finite temperature $T$.
It is believed that the energy required grows linearly with the distance
between the color charge and anticharge due to the formation of a color
electric flux tube. The idea is that a charge or anticharge is a source or
sink of color electric flux, which is the analog of ordinary electric flux
for the strong interactions. But unlike ordinary electric flux, the color electric
flux is expelled from the vacuum and is trapped in a thin flux tube
connecting the color charge and anticharge.
This is very similar to the way that a superconductor expels magnetic
flux and traps it in thin tubes called Abrikosov-Gorkov vortex lines.\\
There is a general statement that the color confinement is
supported by the idea that the vacuum of quantum
Y-M theory is realized by a condensate of monopole-antimonople
pairs [1]. In such a vacuum the interacting field between two
colored sources located in $\vec{x}_{1}$ and $\vec{x}_{2}$ is
squeezed into a tube whose energy $E_{tube}\sim
\vert\vec{x}_{1}-\vec{x}_{2}\vert$. This is a complete dual
analogy to the magnetic monopole confinement in the Type II
superconductor. Since there is no monopoles as classical
solutions with finite energy in a pure Y-M theory, it has been
suggested by 't Hooft [2] to go into the Abelian projection
where the gauge group SU(2) is broken by a suitable gauge
condition to its (may be maximal) Abelian subgroup U(1).
It is proposed that the interplay between a quark-antiquark pair
is analogous to the interaction between a monopole-antimonopole
pair in a superconductor.

In fact, the topology of Y-M SU(N) manifold and that of its
Abelian subgroup $[U(1)]^{N-1}$ are different, and since any such gauge
is singular, one might introduce the string by performing the
singular gauge transformation with an Abelian gauge field
$A_{\mu}$ [3]
\begin{eqnarray}
\label{e1}
A_{\mu}(x)\rightarrow
A_{\mu}(x)+\frac{g}{4\pi}\,\partial_{\mu}\Omega(x)\ ,
\end{eqnarray}
where $\Omega(x)$ is the angle subtended by the closed
space-like curve described by the string at any point
$x=(x^0,x^1)$, and $g=2\pi/e$ is responsible for the magnetic flux
inside the string, $e$ being the Y-M coupling constant. Here, we
choose
a single string in the
two-dimensional  world sheet $y_{x}(\tau,\sigma)$, for
simplicity. Obviously, the Abelian field-strength tensor
$F_{\mu\nu}^{A}=\partial_{\mu}A_{\nu}-\partial_{\nu}A_{\mu}$
transforms as
$$ F_{\mu\nu}^{A}(x)\rightarrow
F_{\mu\nu}^{A}(x)+\tilde {G}_{\mu\nu}(x)\ ,$$
where a new term (the Dirac strength string tensor)
$$ \tilde
{G}_{\mu\nu}(x)=\frac{g}{4\pi}\,\lbrack\partial_{\mu},\partial_{\nu}\rbrack \,
\Omega(x)\ , $$
is valid on the world sheet only [4]
$$ \tilde {G}_{\mu\nu}(x)=\frac{g}{2}\,\epsilon_{\mu\nu\alpha\beta}\,\int\int d\sigma\,
d\tau\,
\frac{\partial (y^{\alpha},y^{\beta})}{\partial (\sigma
,\tau)}\,\,
\delta_{4} \lbrack x-y(\sigma ,\tau)\rbrack \ .$$
Formally, a gauge group element, which transforms a generic SU(N)
connection onto the gauge fixing surface in the space of
connections, is not regular everywhere in space-time. The
projected (or transformed) connections contain topological
singularities (or defects). Such a singular transformation
(\ref{e1}) may form the worldline(s) of magnetic monopoles.
Hence, this singularity leads to the monopole current
$J_{\mu}^{mon}$. This is a natural way of the transformation
from the Y-M theory to a model dealing with Abelian fields.
A dual string is nothing but a formal idealization of a magnetic
flux tube in the equilibrium against the pressure of
surrounding superfluid (the scalar Higgs-like field) which it displaces
[5,6].

The lattice results, e.g., [7] give the promised picture that the
monopole degrees of freedom can indeed form a condensate responsible for the confinement.
By lattice QCD simulations, it is observed that large monopole
clustering covers the entire physical vacuum in the confinement phase,
which is identified as a signal of monopole condensation being responsible
for confinement. The expression for the static heavy quark potential, using an
effective dual Ginzburg-Landau model [8], has been presented in
[9]. In the paper [10], an analytic approximation to the dual
field propagator without sources and in the presence of quark
sources, and an expression for the static quark-antiquark
potential were established.

The aim of this paper is to consider the phase transition in the
"hot" four-dimensional model based on
the dual description of a long-distance Y-M theory which
shows some kind of confinement. We study the model of Lagrangian
where the fundamental variables are an octet of dual potentials
coupled minimally to three octets of monopole (Higgs-like) fields [11].

In the scheme presented in this work, the flux distribution in
the tubes formed between two heavy color charges is understood
via the following statement: the Abelian Higgs-like monopoles are excluded
from the string region while the Abelian electric flux is
squeezed into the string region.

In the model there are the
dual gauge field $\hat{C}_{\mu}^{a}(x)$ and the scalar field
$\hat{B}_{i}^{a}(x)$ ($i=1,..., N_{c}(N_{c}-1)/2$; $a$=1,...,8
 is a color index) which
are relevant modes for infrared behaviour. The local coupling of
the $\hat{B}_{i}$-field to the $\hat{C}_{\mu}$-field provides
the mass of the dual field and, hence, a dual Meissner effect.
 Although
$\hat{C}_{\mu}(x)$ is invariant under the local transformation
of $U(1)^{N_{c}-1}\subset SU(N_{c})$,
$\hat{C}_{\mu}=\vec{C}_{\mu}\cdot\vec{H}$ is an $SU(N_{c})$-gauge
dependent object and does not appear in the real world alone
($N_{c}$ is the number of colors and $\vec{H}$ stands for the
Cartan superalgebra). The scope of commutation relations, two-pint
Wightman functions and
Green's functions as well-defined distributions in the space
$S(\Re^{d})$ of complex Schwartz test functions on $\Re^{d}$,
the monopole- and dual gauge-field propagations, the asymptotic
transverse behaviour of both the dual gauge field and the
color-electric field, the analytic expression for the static
potential can be found in [11].

\section{The string-like flux tube phase transition}

Phase transitions in dual models are associated with a change in symmetry or
more correctly these transitions are related with the breaking of symmetry.
As a starting point we assume, for simplicity, that the model is characterized
by the scalar order-parameter $\langle\hat{B}_{i}(x)\rangle$ = $\hat{B}_{0}$
for the scalar field $\hat{B}_{i}(x)$ identified in the dual model
as the Higgs-like field. The classical partition function looks like
\begin{eqnarray}
\label{e2}
Z_{cl}=\int D\hat{B}_{i}\exp\left\{-\int^{\beta}_{0} d\tau \int d^{3}\vec{x}
L(\tau,\vec{x})\right\} ,
\end{eqnarray}
where $\beta=T^{-1}$.
The dual description of the Y-M theory is simply understood by
switching on the dual gauge field $\hat{C}_{\mu}(x)$ and the
three
scalar octets  $\hat{B}_{i}(x)$ (necessary to give mass to all $C_{\mu}^{a}$
and carrying color magnetic charge) in
the Lagrangian density (LD) $L$ [12]
\begin{eqnarray}
\label{e3}
L=2\,Tr\left [
-\frac{1}{4}\hat{F}^{\mu\nu}\hat{F}_{\mu\nu}+\frac{1}{2}\left
(D_{\mu}\hat{B}_{i}\right )^2\right ] - W\left
(\hat{B}_{i}\right )\ ,
\end{eqnarray}
where
$$ \hat{F}_{\mu\nu}=\partial_{\mu}\hat{C}_{\nu}-\partial_{\nu}\hat{C}_{\mu}-
ig\lbrack\hat{C}_{\mu},\hat{C}_{\nu}\rbrack\ ,$$
$$D_{\mu}\hat{B}_{i}=\partial_{\mu}\hat{B}_{i}-ig\,\lbrack\hat{C}_{\mu},\hat{B}_
{i}\rbrack\ ,$$
The Higgs-like fields
develop their vacuum expectation values (v.e.v.)
$\hat{B}_{{0}_{i}}$ and the Higgs potential $W(\hat{B}_{i})$
has a minimum at $\hat{B}_{{0}_{i}}$ of the order O (100 MeV) defined by the
string tension and the Higgs-like potential $W(\hat{B}_{i})$ has a minimum at
$\hat{B}_{{0}_{i}}$. The v.e.v.
$\hat{B}_{{0}_{i}}$ produce a color monopole generating current
confining the electric color flux [11]:
$$J_{\mu}^{mon}(x) = \frac{2}{3}\partial^{\nu}G_{\mu\nu}(x),$$
where
\begin{eqnarray}
\label{e44}
G_{\mu\nu} = \partial_{\mu}C_{\nu}-\partial_{\nu}C_{\mu} + \tilde G_{\mu\nu},
\end{eqnarray}
$\hat C_{\mu} =\lambda ^{8}C_{\mu}$ ($\lambda^{a}$ is the generator of SU(3)).
The interaction of the dual field with other fields (scalar nonobservable fields)
is due to monopole current $J_{\mu}^{mon}(x)$ in the Higgs-like condensate
$(\chi + B_{0})$ in terms of the dual gauge coupling $g$ up to divergence of the local
fase of the Higgs-like field, $\partial_{\mu}f(x)$:
$$g\,C_{\mu}(x)= \frac{J_{\mu}^{mon}(x)}{4\,g\,(\chi + B_{0})^{2}} +\partial_{\mu}f(x).$$

As a result, we obtained [11] that the dual gauge field is defined by the
divergence of the Dirac string tensor $\tilde G_{\mu\nu}$ shifted by the
divergence of the scalar Higgs-like field. For large enough $\vec x$, the monopole
field is going to its v.e.v., while $C_{\mu}(\vec x\rightarrow\infty)\rightarrow 0$
and $J_{\mu}^{mon}(\vec x\rightarrow\infty)\rightarrow 8\,m^2\,C_{\mu}$ with
$m$ being the mass of $C_{\mu}$ field. The Higgs-like fields are associated with not
individual particles but the subsidiary objects in the massive gauge theory. These fields
cannot be experimentally observed as individual particles.\\
 It is believed that LD
(\ref{e3}) can generate classical equations of motion
carrying a unit of the $z_{3}$ flux confined in a narrow tube along
the $z$-axis (corresponding to quark sources at
$z=\pm\infty$). This is a dual analogy to the Abrikosov [13]
magnetic vortex solution. \\
The question is   what happens with the flux tube in excited matter at nonzero
temperature $T$. At $T\neq 0$ the oscillations of the flux tube become visible
up to the energy of excitation $e_{\beta} = s/\beta$ at certain entropy density
$s$.
In Eq. (\ref{e2}) the sum is taken over fields $\hat{B}_{i}$ periodic in
Euclidean time $\tau$ with the period $\beta$, and the thermal equilibrium
in flat space-time at temperature $T$ is considered. The scalar
Higgs-like field is only part of the dual picture, however it is the only field
that we shall make visible in the path integral (\ref{e2}).
The Fourier expanding $\hat{B}(\tau,\vec{x})$ as
\begin{eqnarray}
\label{e4}
\hat{B}(\tau,\vec{x})=\hat{B}(\vec{x}) + \sum_{n=1}\hat{B}_{n}(\vec{x})
\exp (2\pi i n\tau/\beta)
\end{eqnarray}
can allow one to reflect the periodicity of $\hat{B}(\tau,\vec{x})$ in
imaginary time. Note, that the first term in (\ref{e4}) gives the
zero-temperature mode while the other ones count the "heavy"
high-temperature modes.\\
We now move to a simple physical pattern: let us define the "large" and "small" systems.
It is known that in classical mechanics, the stochastic processes in a dynamic
system are under the weak action of a "large" system. "Small" and "large" systems
are understood to mean that the number of the states of freedom of the former is
less that that of the later. The"large" system is supposed to be in the equilibrium state
(thermostat with the temperature $T$). We do not exclude the interplay between two systems.
The role of the "small" system is played by
the restricted region of confined charges, the flux tube that we study in this paper.
The stationary stochastic processes in the deconfined state are distorted by the random
source $\tilde G_{\mu\nu}(x)$ in the dual field tensor $G_{\mu\nu}(x)$ (\ref{e44}),
and under the weak action of a "large"
system described by the scalar field $\phi$ in the Lagrangian density term
${\vert(\partial_{\mu} - i\,g\,C_{\mu})\phi\vert}^{2}$ [11].

As a result, in the dual Higgs model [11] the finite energy of the peace of
the isolating string-like flux tube of the length $R$ keeps growing as $R$
\begin{eqnarray}
\label{e5}
E(R) \simeq \frac{\xi\,\vec{Q}^{2}_{\alpha}}{16\,\pi}m^{2}
R(12.4 -6\ln{\tilde\mu\,R}),
\end{eqnarray}
where $\vec{Q}_{\alpha}=e\,\vec{\rho}_{\alpha}$ is the Abelian color
electric charge, while $\vec{\rho}_{\alpha}$ is the weight vector of the
SU(3) algebra; $\xi$ is the model
constant coming from the commutator for the dual vector fields; $\tilde\mu$
is the infrared mass parameter.\\
Let us introduce the canonical partition function
\begin{eqnarray}
\label{e6}
Z_{c}=\sum_{flux~tube~configurations} \exp [-\beta\,E (R)] =
\sum_{R} N(R)\exp [-\beta\,E (R)]
\end{eqnarray}
for ensembles of systems with a single static flux tube, where $N(R)$ is
the number of configurations of the flux tube of length $R$. The sum (\ref{e6})
is divergent if $T>T_{0}$ where the limiting temperature $T_{0}$ is defined
taking also into account $N(R)$.
In this form,
$N(R)$ can be considered in the discrete space of a scalar Higgs-like condensate
and one requires
the flux tubes to lie along the links of a 3-dimensional cubic lattice of
volume $V$ with the lattice size $l \sim\mu^{-1} = (\sqrt{2\,\lambda}\,\hat{B}_{0})^{-1}$,
where $\mu$ is the mass of the scalar Higgs-like field and $\lambda$ is its coupling
constant. In the rest of physics, for $l<< R$ the number of configurations
$N(R)$ is interpreted in terms of the entropy density $s$ of the flux tube
by a fundamental formula
\begin{eqnarray}
\label{e7}
\tilde N(R)=e^{\tilde s},
\end{eqnarray}
where $\tilde N(R) = N(R)\,c\,l^{3}/V$, $\tilde s =s\,R/l$,
$c$ is the positive constant of the order $O(1)$ (see also [14]). The relation (\ref{e7})
counts the flux tubes that do not intersect the volume boundary (bound state).
They called "short" flux tubes. If the entropy density $s$, which
was inferred from classical reasoning, is like every other entropy density that
we have met, then a flux tube has a very large number of configurations, roughly
$N(R)\sim\exp [s(R/l)]$. Can one by some sort of calculation count the number
of configurations of flux tube and reproduce the formula (\ref{e7}) for the entropy
density? For this, we need a quantum theory of confinement, so, at present at least,
dual Y-M theory is the only candidate. Even in this theory, the question was out of
reach for last three decades.\\

Inserting (\ref{e7}) into Eq. (\ref{e6}) one gets
\begin{eqnarray}
\label{e8}
Z_{c}=\frac{V}{l^{3}} \sum_{R} \exp [-\beta\,\sigma_{eff}(\beta)\,R],
\end{eqnarray}
where
\begin{eqnarray}
\label{e9}
\sigma_{eff}(\beta)=\tilde\sigma_{0}-\frac{s}{l\,\beta}
\end{eqnarray}
is the order parameter of the phase transition
and
\begin{eqnarray}
\label{e10}
\tilde\sigma_{0}=\sigma_{0}\left (1-\frac{1}{4}\ln\frac{\tilde\mu^{2}}{m^{2}_{R}} \right)\,,\,\,
\sigma_{0}=\frac{3}{4}\alpha (Q)\,m^{2}=\frac{3}{4}\frac{\pi}{g^{2}}m^{2}
\end{eqnarray}
with  $\alpha (Q)$ being the running coupling constant.
Here, $R$ in the logarithmic function in (\ref{e5}) has been
replaced by the characteristic length $R_{c}\sim 1/m_{R}$ which
determines the transverse dimension of the dual field
concentration, while $\tilde{\mu}$ is associated with the inverse
"coherent length" and the dual field mass $m$ defines
the "penetration depth" in the Type II superconductor where $m<\mu$.
We got $\sigma_{0}\simeq 0.18~GeV^2$ [11] for the mass of the dual
$C_{\mu}$-field $m=0.85~GeV$ and $\alpha =e^2/(4\,\pi)$=0.37
obtained from fitting the heavy quark-antiquark pair spectrum
[15]. The value of the string tension (\ref{e9}) at $T=0$ is
close to a phenomenological one (e.g., coming from the Regge
slope of the hadrons).
Making the
formal comparison of the result obtained here in the analytic form,
we recall the  expression of the
energy per unit length of the vortex in the Type II superconductor
[16,9]
\begin{eqnarray}
\label{e11}
\epsilon_{1}=\frac{{\phi_{0}}^2\,m_{A}^2}{32\,\pi^{2}}\,\ln\left
(\frac{m_{\phi}}{m_{A}}\right )^2\ ,
\end{eqnarray}
where $\phi_{0}$ is the magnetic flux of the vortex, $m_{A}$ and
$m_{\phi}$ are penetration depth mass and the inverse coherent length,
respectively. On the other hand, the string tension in
Nambu's paper (see the first ref. in [1]) is given by
\begin{eqnarray}
\label{e12}
\epsilon_{2}=\frac{g_{m}^2\,m_{v}^2}{8\,\pi}\,\ln\left
(1+\frac{{m_{s}}^2}{{m_{v}}^2}\right )\ ,
\end{eqnarray}
with $m_{s}$ and $m_{v}$ being the masses of scalar and vector
fields and $g_{m}$ is a magnetic-type charge. It is clear that for a
sufficiently long string $R >>m^{-1}$ the $\sim R$-behaviour of
the static potential is dominant; for a short string $R<<m^{-1}$
the singular interaction provided by the second term in (\ref{e5})
becomes important if the average size of the monopole is even
smaller.

Our description is characterized by a limiting temperature $T_{0}>T$, and it is evident that
the flux tube picture cannot work at $T > T_{0}$, where
\begin{eqnarray}
\label{e15}
T_{0}=\frac{3}{4}\frac{1}{s}\alpha(Q)\frac{m^2}{\mu}
\left (1-\frac{1}{4}\ln\frac{\tilde\mu^{2}}{m^{2}_{R}} \right)
\end{eqnarray}
for which $\sigma_{eff}(T_{0}) = 0$.
The vacuum expectation value $B_{0}$ is the threshold energy to excite the
monopole (Higgs-like field) in the "QCD" vacuum. It corresponds to the
Bogolyubov particle in the ordinary superconductor. In case of such excitations
exist, the phase transition is expected to occur at $T_{0}\sim $ 200 MeV. The
value $B_{0}\simeq$ 276 MeV is regarded as the ultraviolet cutoff of the theory.
At sufficiently high temperature QCD definitely loses
confinement and the flux tube definitely disappears.
It is evident that at $T\rightarrow T_{0}$ the
flux tube becomes arbitrary long.
As a result, the temperature-dependent mass $m(\beta)$ of the dual gauge field
$\hat{C}_{\mu}$ is derived as follows
\begin{eqnarray}
\label{e16}
m^{2}(\beta) = \frac{4}{3}\frac{\sigma_{eff}(\beta)}{\alpha (Q,\beta)}.
\end{eqnarray}
Obviously, $m(\beta)\rightarrow m$ as $\beta\rightarrow\infty$, and
$m(\beta)\rightarrow 0$ as $1/\beta\rightarrow T_{0}$. The latter limit means that
\begin{eqnarray}
\label{e17}
\frac{1}{3}\partial^{\nu} \tilde G_{\mu\nu}= (m^{2}\,C_{\mu} + 4\,m\,\partial_{\mu} \bar b)\rightarrow 0
\end{eqnarray}
as $T\rightarrow T_{0}$ (here, $\bar b$ is the Higgs-like field). On the other hand, the divergence of
$\tilde G_{\mu\nu}$ is just the current carried by a charge $g$ moving along the path $\Gamma$:
\begin{eqnarray}
\label{e18}
\partial^{\nu} \tilde G_{\mu\nu}(x)= - g\,\int_{\Gamma} dz_{\mu}\,\delta^{4}(x-z).
\end{eqnarray}
Hence, $\partial^{\nu} \tilde G_{\mu\nu}(x)\rightarrow 0$ as $g\rightarrow 0$. And the final remark
concerning the zeroth value of $m(\beta)$: recall that $m^{2}(\beta)\sim g^{2}\,\delta^{2}(0)$, where
$\delta^{2}(0)$ is the inverse cross section of the flux tube. This cross section is infinitely large
if $m\rightarrow 0$.
Actually, $\sigma_{eff}(\beta)$ is the effective measure of the phase
transition when the flux tube is produced. The fact that $\sigma_{eff}(\beta_{0})=0$
means the special phase where two color charges are separated from each other by
infinite distance.
At $T=T_{0}$ the entropy and the total energy are related to each other by $s=E/T_{0}$. The
level density of a system is $e^{s}$, therefore $s=E/T_{0}$ implies an exponentially
rising mass spectrum if one identifies $E$ with the mass of a quark-antiquark bound state. \\
It is assumed that the flux tube starts in a thermal exciting phase, a phase in which
the flux tube is quasi-static and in thermal equilibrium at a temperature close to
$T_{0}$, the limiting temperature of the dual model. We assume that the string coupling
is sufficiently small and the local space-time geometry is close to the flat over
the length scale of the finite size box-block of volume $v=r^{3}$.
In each block $j$ the flux tube is homogeneous and isotropic with the energy $E_{j}$.
It was shown [17] that applying the string thermodynamics
to a volume $v=r^{3}$ in the string gas cosmology one can get the mean square mass fluctuation
in a region of radius $r$ evaluated at the temperature close to $T_{0}$:.
\begin{eqnarray}
\label{e17}
\langle (\delta \mu)^{2}\rangle = \frac{r^{2}}{R^{3}}\frac{1}{\beta - \beta_{0}},
\beta > \beta_{0}.
\end{eqnarray}
What is important that the heat bath turns out to scale as $r^2$. This is typical
string-like effect while in the case of the point-like object thermodynamics, the
result would scale as $r^3$.

\section{$T$-dependent coupling constant}
To make sense of real processes at high energies in a hot environment
one has to define a parameter of perturbative expansion $\alpha$ as a function of $T$.
We make the formal replacement $\alpha (Q) \rightarrow \alpha (Q,T)$ by a formula
\begin{eqnarray}
\label{e15}
\alpha^{-1}(Q,T)\simeq b\ln[Q/\Lambda (T)] = b\{\ln(Q/\Lambda) - \ln[\Lambda (T)/\Lambda]\},
\end{eqnarray}
where $b$ is the well-known coefficient depending on the number of flavors. Actually, we
define $\ln[\Lambda (T)/\Lambda]$ through the one-loop finite temperature renormalization
constant $z^{(1)}(T)$ as
\begin{eqnarray}
\label{e16}
z^{(1)}(T)=-b\ln[\Lambda (T)/\Lambda].
\end{eqnarray}
Then
\begin{eqnarray}
\label{e17}
\alpha^{-1}(Q,T)\simeq \alpha^{-1}(Q)+ z^{(1)}(T)
\end{eqnarray}
and
\begin{eqnarray}
\label{e18}
\alpha(Q,T)= [1 + \rho(T)]\alpha(Q),
\end{eqnarray}
where
\begin{eqnarray}
\label{e19}
z^{-1}(T) = 1 + \rho(T)=\{1- b\ln[\Lambda (T)/\Lambda]\alpha(Q)\}^{-1}.
\end{eqnarray}
The $T$-dependent $\Lambda$ is defined by [18]
\begin{eqnarray}
\label{e20}
\ln\frac{\Lambda(T)}{\Lambda} = -\frac{1}{b}\int^{\kappa=T/Q}_{0}
y(x)\frac{dx}{x},
\end{eqnarray}
where $y(x)$ is the leading one-loop coefficient in renormalization
group equation
\begin{eqnarray}
\label{e21}
\frac{\partial \alpha(Q,T)}{\partial\kappa}=\bar\beta_{\kappa}(\alpha,T)
\simeq -y(T)\alpha^{2}(Q).
\end{eqnarray}
In fact, the effective coupling $\alpha(Q,T)$ renormalized at nonzero
temperature depends on the type of vertices in QCD. This might cause
troubles in carrying out the calculations because of the problem related to
thermalized equilibrium within an environment. Based on the analyzes
done in [18] for $\Lambda(T)$ in the case of fermion-gluon
($fg$) and three gluon ($3g$) vertices one can conclude that $\ln[\Lambda(T)/\Lambda]$
decreases as a quadratic function of $T$ and increases as a linear
function of $T$, respectively.

In gauge theories at $T\neq 0$ thermal fluctuations of the gluon act to screen
the electric field component of the gluon, through the development of
temperature-dependent electric mass $m_{el}\sim g\, T$. Recent studies show remarkable facts that
instantons are related to monopoles in the Abelian gauge although these
topological objects belong to different homotopy group. It is known that both
analytical and lattice studies can show a strong correlation between instantons and
monopoles in the Abelian projected theory of QCD. It can be  postulated that at
finite $T$ the running coupling would be replaced by the static screened charge
\begin{eqnarray}
\label{e23}
\frac{1}{\alpha(Q,T)} = \frac{1}{\alpha(Q)}\left \{\left [1-
\frac{\Pi^{00} (q^{0}=0,\vec q\rightarrow 0;\beta)}{\vec Q^{2}}\right ]
+\frac{\alpha (Q)}{6\,\pi}\left (\frac {11\,N}{2} - N_{f}\right )\,\ln\frac{\vec{Q}^{2}}
{M^{2}}\right \},
\end{eqnarray}
for gauge group $SU(N)$,
where $M$ is the renormalization energy scale and the inverse screening length is
given by the gluon self-energy $\Pi^{\mu\nu}(q)$ at the lowest order of $g^{2}$
in hot theory containing the quark fields with the mass $m_{q}$ (see, e.g., [19])
\begin{eqnarray}
\label{e24}
- \Pi^{00} (q^{0}=0,\vec q\rightarrow 0;\beta) = g^{2}\,T^{2}
\left [\frac{N}{3} + \frac{N_{F}}{\pi^2\,T^{2}}\,I_{F}(\beta, \bar\mu, m_{q}) \right ]= m^{2}_{el}(\beta),
\end{eqnarray}
where
\begin{eqnarray}
\label{e25}
I_{F}= \int^{\infty}_{0}\,\frac{dx\, x^2}{\sqrt{x^2 + m^{2}_{q}}} [n_{F}(x^2) + \bar n_{F}(x^2)],
\end{eqnarray}
\begin{eqnarray}
\label{e26}
n_{F}(x^2)= \frac{1}{\exp[(\sqrt{x^2 + m^{2}_{q}}-\bar\mu)\beta]+1},\,\,\,
\bar n_{F}(x^2)= \frac{1}{\exp[(\sqrt{x^2 + m^{2}_{q}}+\bar\mu)\beta]+1},
\end{eqnarray}
$N_{F}$ is the number of quarks, the chemical potential $\bar\mu$ is defined by
the baryon density $\rho$ in the formula
\begin{eqnarray}
\label{e27}
\rho\sim\int\frac{d^{3}x}{(2\,\pi)^3}[n_{F}(x^2)-\bar n_{F}(x^2)].
\end{eqnarray}
The first term in (\ref{e24}) refers to pure $SU(N)$ gauge theory.
 Hence, $\alpha^{-1}(Q,T)$
has the following expansion over $\vec Q^{2}/M^{2}$ and $T^{2}/\vec Q^{2}$:
\begin{eqnarray}
\label{e28}
\frac{1}{\alpha (Q,T)}= \frac{1}{\alpha (Q)} + \frac{1}{6\,\pi}\left (\frac{11\,N}{2}
- N_{f}\right )\,\ln\frac{\vec Q^{2}}{M^{2}} + 4\,\pi\frac{T^{2}}{\vec Q^{2}}
\left [\frac{N}{3} +\frac{N_{F}}{\pi^2\,T^{2}}\,I_{F}(\beta, \bar\mu, m_{q})\right],
\end{eqnarray}
where $\alpha (Q,T)\rightarrow 0$ as $T\rightarrow\infty$.

Because quark confinement is considered here as the dual version of the confinement
of magnetic point charges in Type-II superconductor (magnetic Abrikosov vorteces), the
upper limit for $T_{0}$ is given by the requirement $(m/\mu) < 1$, i.e.,
\begin{eqnarray}
\label{e29}
T_{0} < \frac{3}{4}\,\alpha(Q)\,m\left ( 1 -\frac{1}{4}\ln\frac{\tilde\mu^{2}}{m^{2}_{R}}\right ).
\end{eqnarray}
Numerical estimation leads to $ T_{0} < 222$ MeV at $B_{0}\simeq$ 276 MeV,
$\alpha = 0.37$ and $ m= 0.85 $ GeV [11] for $m_{R} \sim \tilde\mu$  and $s\sim O(1)$.
Note that above estimated upper limit on $ T_{0}$ is less than the typical value $T_{c}\simeq 256$ MeV
of the critical temperature of the quark-gluon plasma phase transition (see, e.g., [20]).

\section{Flux tube solutions}

The temperature-dependent flux-tube solution for the dual gauge filed along the
z-axis (within the cylindrical symmetry) has the following asymptotic transverse
behaviour (for details see [11] at $T$=0)
\begin{eqnarray}
\label{e30}
\tilde C(r,\beta)\simeq \frac{4\,n}{7\,g(\beta)}-\sqrt{\frac{\pi\,m(\beta)\,r}
{2\,\kappa}}\,e^{-\kappa m(\beta)r}\left [1+\frac{3}{8\,\kappa\,m(\beta)\,r}\right],
\end{eqnarray}
where $r$ is the radial coordinate (the distance from the center of the flux-tube), $n$ is
the integer number associated with the topological charge [21], $\kappa = \sqrt{21}$.

The color-electric field $E$ inside the quark-antiquark bound state is given by the
rotation of the dual gauge field
\begin{eqnarray}
\label{e31}
\vec{E}=\vec{\nabla}\times\vec{C}=\frac{1}{r}\frac{d\tilde {C}(r)}{dr}\vec{e}_{z}\simeq
E_{z}(r)\cdot\vec{e}_{z},
\end{eqnarray}
where $\vec{e}_{z}$ is a unit vector along the $z$-axis, and the $T$-dependent $E_{z}(r,\beta)$
looks like [11]
\begin{eqnarray}
\label{e32}
E_{z}(r,\beta)= \sqrt{\frac{\pi\,m(\beta)}{2\,\kappa\,r}}\,e^{-\kappa m(\beta)r}
\left [\kappa\,m(\beta) - \frac{1}{2\,r}\right].
\end{eqnarray}
The lower bound on $r=r_{0}$ can be estimated from the relation $r_{0} > [2\,\kappa\,m(\beta)]^{-1}$ which
leads to $r_{0}>0.03$ fm at $T=0$. Obviously, $r_{0}\rightarrow\infty$ as $m(\beta)\rightarrow 0$ at
$T\rightarrow T_{0}$ (deconfinement).

In Fig. 1, we show the dependence of $m$ as a function of the temperature $T$ at
different scale parameters $M$. No dependence found on quark current masses (we used
$m_{q}$ = 7, 10 and 135 MeV). No essential dependence found for different $N_{f}$ and $N_{F}$.


\begin{figure*}[h!]
\begin{center}
\mbox{\epsfig{figure=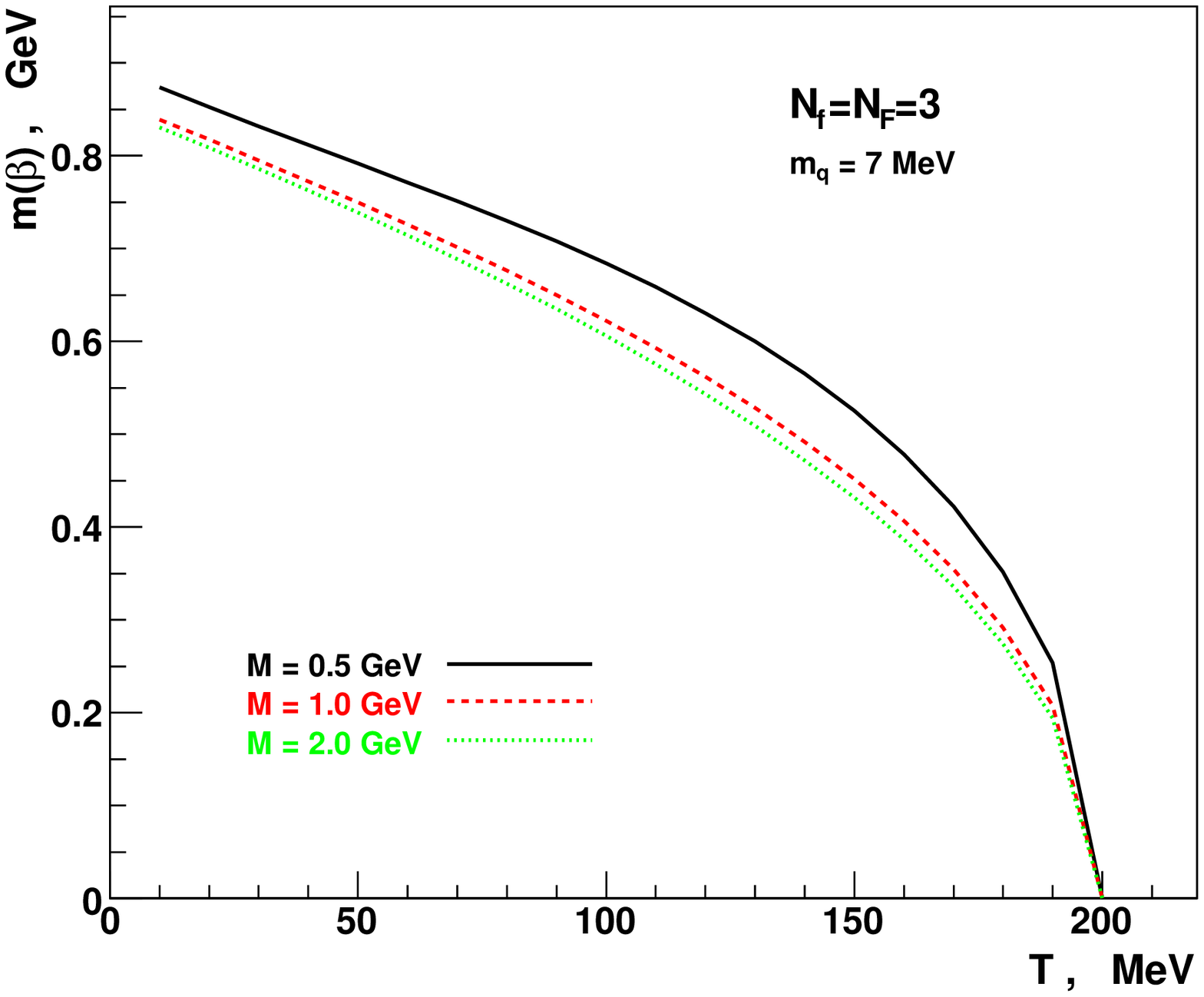,
width=0.48\textwidth,height=0.325\textheight}} \hfill
\mbox{\epsfig{figure=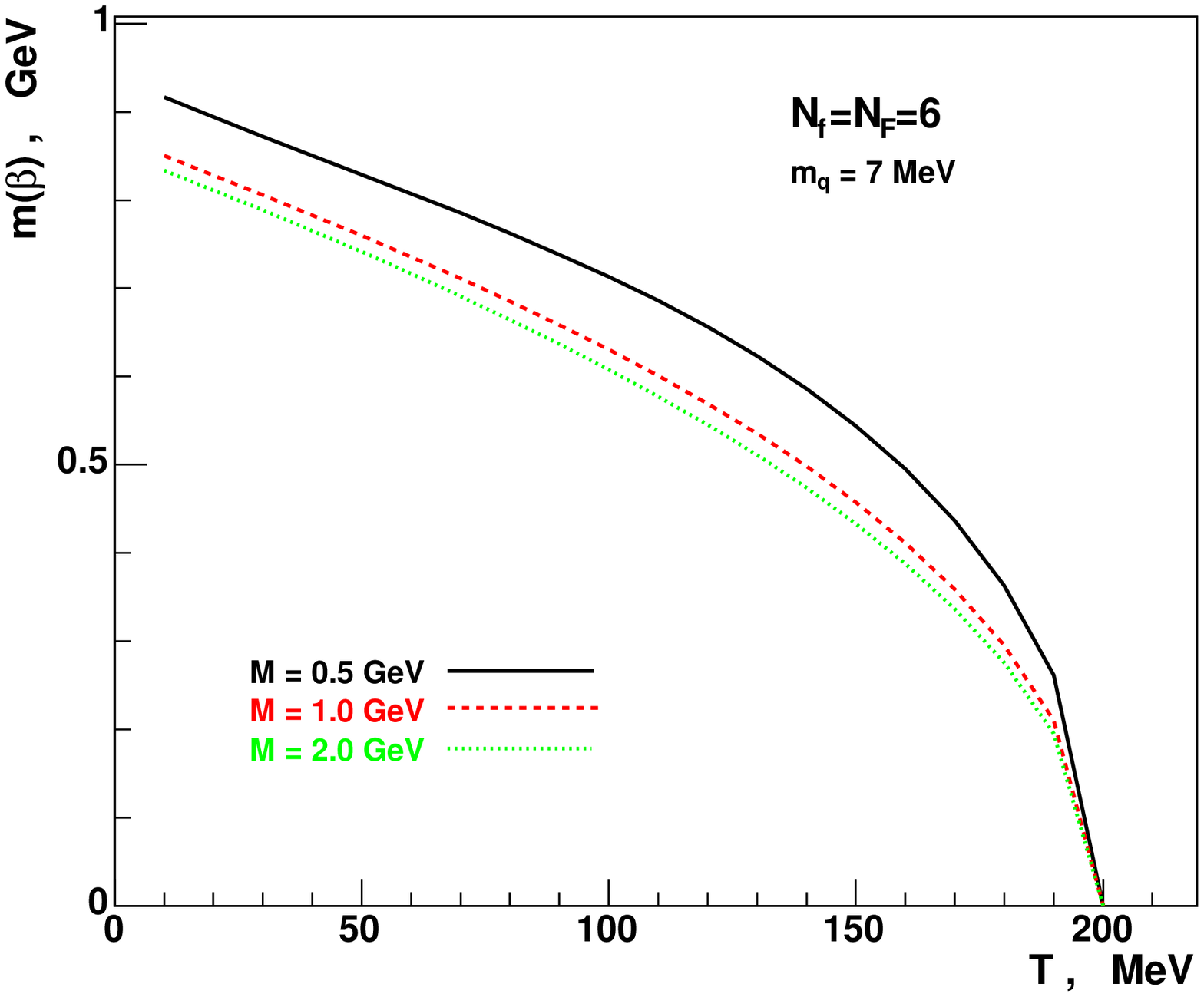,width=0.48\textwidth,height=0.325\textheight}}
\end{center}
 \vspace*{-5mm}
\caption{The dual gauge boson mass $m(\beta)$ shown as a funciton
of $T$ at different renormalization energy scale $M$ and fixed
value $m_{q} = 7$ MeV with a) $N_{f} = N_{F} = 3$ b) $N_{f} = N_{F} =
6$ \label{mass}}
\end{figure*}

In Fig. 2 and Fig. 3, we show numerical solutions of the flux
tube, namely, the profiles of the transverse behaviour of $\tilde
C (r,\beta)$ and the color electric field $E_{z}(r,\beta)$,
respectively, as functions of radial variable $r$ at different
temperatures. We found rather sharp increasing of $\tilde C
(r,\beta)$ at small values of $r$. No essential
dependence of $r$ emerges in the region $r> 0.1$ fm. The field
$E(r,\beta)$ dissapperas when the temperature close to $T_{0}$.

\begin{figure*}[h!]
\begin{center}
  \mbox{\epsfig{figure=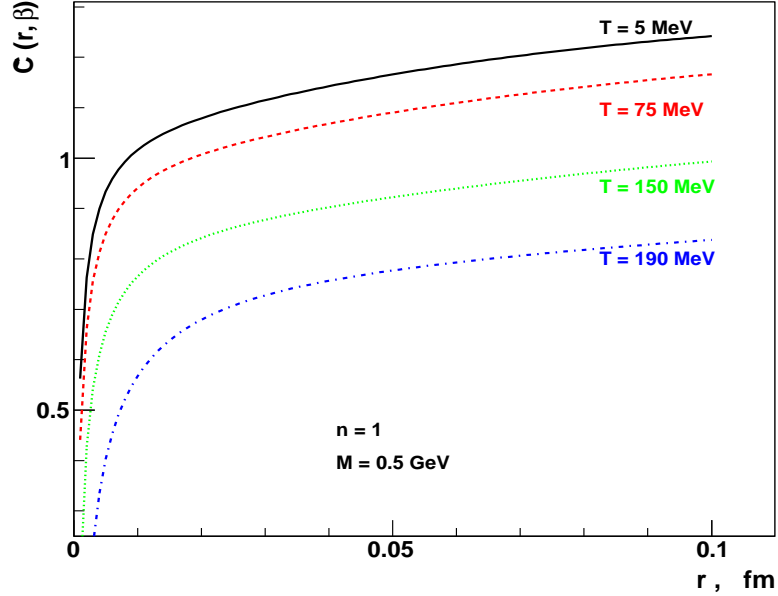, width=0.65\textwidth,height=0.38\textheight}}
\end{center}
\vspace*{-5mm} \caption{The profiles of the dual gauge boson
field $\tilde C (r,\beta)$ shown as a funciton of the radial
coordinate $r$ at different $T$. \label{gauge}}
\end{figure*}


\begin{figure*}[h!]
\begin{center}
  \mbox{\epsfig{figure=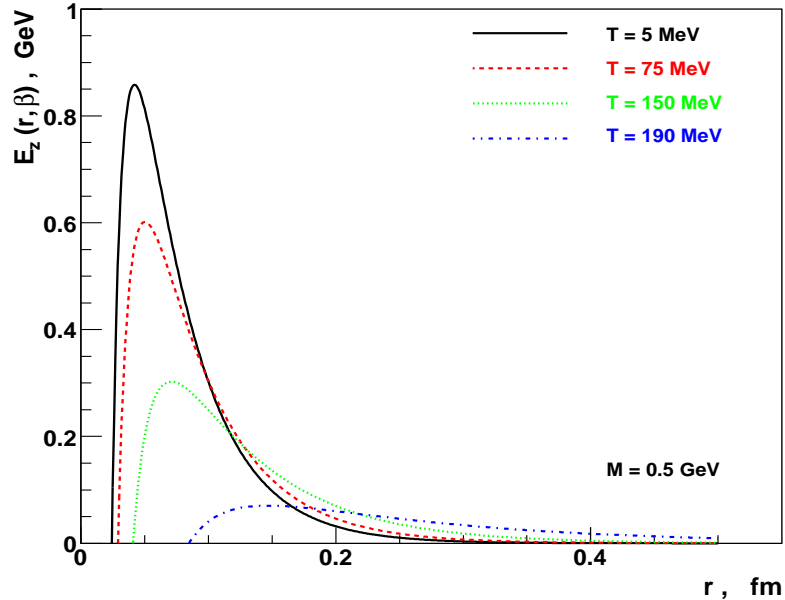, width=0.65\textwidth,height=0.38\textheight}}
\end{center}
\vspace*{-5mm} \caption{The profiles of the color-electric
field $E(r,\beta)$ as a funciton of the radial coordinate $r$ at
different $T$. \label{energy}}
\end{figure*}


\section{Summary}

We were based on the dual gauge model of the long-distance Yang-Mills
theory in terms of two-point Wightman functions.
 Among the
physicists dealing with the models of interplay of a scalar
(Higgs-like) field with a dual vector (gauge) boson field,
where the vacuum state of the quantum Y-M theory is realized by a
condensate of the monopole-antimonopole pairs, there is a strong
belief that the flux-tube solution explains the scenarios of
color confinement. Based on the flux-tube scheme approach of
Abelian dominance and monopole condensation, we obtained the
analytic expressions for both the monopole and  dual gauge
boson field propagators [11]. These propagators lead to
a consistent perturbative expansion of Green's functions.

The monopole condensation, formulated in the
framework of long-distance Y-M model, causes the strong and long-range
interplay between heavy quark and antiquark, which gives the
confining force, through the dual Higgs mechanism.
The analytic expression for the static potential at large distances
grows linearly with the distance $R$ apart from
logarithmic correction.

Making an analytic comparison of $\epsilon_{1}$ (\ref{e11})
and $\epsilon_{2}$ (\ref{e12}) with $\sigma_{eff}$ at $T=0$ in (\ref{e9}),
one can conclude that we have obtained a similar behaviour of
the string tension $\sigma_{eff}$ to those in the magnetic flux picture of
the vortex and in the Nambu scheme, respectively, as well as
in the dual Ginzburg-Landau model [9].

We observed that the flux tube can be produced abundantly when the
phase transition emerges at the temperature $T=T_{0}$, obeying the
condition $\sigma_{eff}(T=T_{0}) = 0$.
We found that the phase transition temperature essentially depends
on $\alpha (Q)$ and the mass of the dual gauge field $m$.

We have presented this subject if one plays the game with the
choice of the gauge group where the Abelian group appears as a
subgroup of the full Y-M gauge group. This is a very useful
method of calculating the confinement potential in the static
limit in the analytic form. However, we understand that no real
physics can depend on a choice of gauge group. We conclude with some
next steps [22] in
more formal consideration of the Y-M theory where it seems to be a
new mechanism of confinement.



\end{document}